\documentstyle[prl,aps,multicol,psfig]{revtex}
\begin{document}
\draft

\title{Critical properties of the reaction -- diffusion model
2A $\rightarrow$ 3A, 2A $\rightarrow$ $\emptyset$}

\author{Enrico Carlon$^{(1,2)}$, Malte Henkel$^{(1)}$ and
Ulrich Schollw\"ock$^{(3)}$}

\address{$^{(1)}$Laboratoire de Physique des Mat\'eriaux,$^{*}$
Universit\'e Henri Poincar\'e Nancy I, \\
B. P. 239, F-54506 Vand{\oe}uvre-les-Nancy Cedex, France}

\address{$^{(2)}$ INFM, Dipartimento di Fisica, Universit\'a di Padova,
I-35131 Padova, Italy}

\address{$^{(2)}$ Sektion Physik, Ludwig - Maximilians -
Universit\"at M\"unchen,
Theresienstr. 37/III, D-80333 M\"unchen, Germany}

\date \today

\maketitle

\begin{abstract}
The steady-state phase diagram of the one-dimensional reaction-diffusion 
model $2A \rightarrow 3A$, $2A \rightarrow\emptyset$ is studied through the 
non-hermitian density matrix renormalization group. In the absence of 
single-particle diffusion the model reduces to the pair-contact process, which 
has a phase transition in the universality class of Directed Percolation (DP) 
and an infinite number of absorbing steady states. 
When single-particle diffusion is added, the number of absorbing steady states 
is reduced to two and the model does not show DP critical behaviour anymore. 
The exponents $\theta=\nu_{\|}/\nu_{\perp}$ and $\beta/\nu_{\perp}$ are 
calculated numerically. 
The value of $\beta/\nu_{\perp}$ is close to the value of the Parity Conserving
universality class, in spite of the absence of local conservation laws. 
\end{abstract}


\begin{multicols}{2} \narrowtext

\section{Introduction}
\label{sec:intr}

Reaction-diffusion systems have attracted considerable interest in 
the past few years \cite{Priv96,Marr99,Hinr00}. 
While at sufficiently high spatial dimensions 
their critical behaviour is correctly described by mean field rate
equations, at dimensions below the upper critical dimension, where
the effect of fluctuations becomes
important this approach is not 
valid anymore. In this case one has to resort to other methods as, 
for instance, field theory \cite{Card96}, exact calculations 
via Bethe ansatz \cite{Bethe-Ansatz}, Monte Carlo and cellular automaton
simulations (see \cite{Marr99,Chop98} and references therein)
or exact diagonalization techniques \cite{Henk90,Mend99}. 
Recently other types of approaches have been proposed as the density 
matrix renormalization group \cite{Carl99} and the 
standard real-space renormalization group \cite{Hooy99}.

In this paper we study the critical properties of a one-dimensional 
reaction-diffusion model where the local dynamics is given by the 
following rules. Consider a single species of particles ($A$) on an 
one-dimensional lattice. Each lattice site can be either occupied by 
a single particle or be empty. A single particle may hop to an empty 
neighbouring site (diffusion). Two particles on neighbouring sites can 
either annihilate ($2A \rightarrow \emptyset$) or give birth to a new particle 
($2A \rightarrow  3A$) in the case that one of the sites next to the couple is 
empty. The reaction rates for these processes are defined in 
equations (\ref{2Ato3A},\ref{2Ato0},\ref{A0to0A}) below. 

Recently, Howard and T\"auber \cite{Howa97} discussed a generalized 
version of the above model where $n$ particles may be created through the
reaction $2A \to (n+2) A$, 
and with an arbitrary number of particles per site. 
They employed a field-theoretical approach and found that the theory is not 
renormalizable. They argued, however, that for all $n$ the
model should not be in the directed percolation (DP) universality class. 
Their conjecture is based on the analysis of the associated master equation 
and on the massless nature of the field theory which is at odds with a 
phase transition of DP type.
However, the critical behaviour of the system, i.e. its universality class,
remained unknown. In absence of diffusion, 
with at most one particle per site, the $n=1$ case 
is the Pair Contact Process (PCP), where the steady-state
phase transition belongs to the DP universality class \cite{Jens93,Dick98}.

Here we present a non-perturbative study of 
the model corresponding to the $n=1$ case of \cite{Howa97}:
the PCP model with additional single-particle diffusion,
as obtained from density matrix renormalization 
group ({\sc dmrg}) calculations. 
The {\sc dmrg} was introduced by White \cite{Whit92} in 1992 
to investigate numerically ground state properties of quantum spin chains. 
Because of its great accuracy, reliability and the possibility of treating 
large systems with a limited computational effort, 
the {\sc dmrg}\ has since been 
applied to an ever increasing set of problems, as reviewed in \cite{DMRG99}. 
In particular, some recent studies were devoted to the application of the
{\sc dmrg}\ to non-hermitian 
problems \cite{Carl99,Burs96,Hiei98,Kaul98,Kaul99,Mars97} 
which appear frequently in many domains of physics, for example in 
low-temperature thermodynamics of spin chains and ladders, for models of 
the non-integer quantum Hall effect and in one-dimensional non-equilibrium 
systems. Reaction-diffusion systems belong to the latter class of models. 
New insight should be expected from the application of the {\sc dmrg}\  to 
them \cite{Carl99}. 

The paper is organized as follows: in Section \ref{sec:mod} we define 
the model and recall some facts about reaction--diffusion systems. 
In Section \ref{sec:shf} we introduce a shift strategy to project out a 
trivial ground state from the quantum Hamiltonian. This improves the 
convergence of the {\sc dmrg}. The method works if an exact expression 
for the eigenstate is available. In Section \ref{sec:pd} we discuss the 
calculation of the critical points and exponents for the model and 
show that the model does not belong to the DP universality class. 
Section \ref{sec:dis} concludes the paper.

\section{The model}
\label{sec:mod}

We consider a one-dimensional lattice of length $L$ with open boundary 
conditions, as usual in {\sc dmrg}\  calculations \cite{DMRG99}. The reactions 
occur with the following rates
\begin{eqnarray} 
\left\{
\begin{array}{ccc}
AA \emptyset & \rightarrow & AAA \\
\emptyset AA & \rightarrow & AAA
\end{array}
\right.
\,\,\,\,\,\,\, &{\rm with \,\,\, rate}& 
\,\,\,\,\,\,\,  \frac{(1-p)(1-d)} 2
\label{2Ato3A} \\
AA~ \rightarrow  ~\emptyset \emptyset~~~~~~~~~~ 
  &{\rm with \,\,\,rate}& \,\,\,\,\,  p(1-d)  
\label{2Ato0}\\
A \emptyset~ \leftrightarrow  ~\emptyset A~~~~~~~~~~
   &{\rm with \,\,\,rate}& \,\,\,\,\,  d  
\label{A0to0A}
\end{eqnarray} 
and are parametrized by the diffusion constant $d$ and the pair annihilation 
rate $p$. This defines the Pair contact process with diffusion (PCPD) model. 

For a first qualitative overview, we consider the mean field kinetic 
equations. To discuss the effects of diffusion we need the kinetic
equations in the {\em pair approximation}, see e.g. \cite{Marr99}. 
If $n=n(t)$ is the spatially averaged single particle density and $c=c(t)$ the 
spatially averaged pair density, we find
\begin{eqnarray} 
\dot{n} &=& -2(1-d)p\, c +(1-d)(1-p)\, (n-c)c/n \\
\dot{c} &=& -(1-d)p\, c\, \frac{2 c+n}{n} 
-2d\,\frac{(n-c)(c-n^2)}{n(1-n)} \nonumber \\ 
 & &+(1-d)(1-p)\,(n-c)(1-c)c/(n(1-n)) 
\end{eqnarray}  

In the $d\rightarrow  1$ limit, non-trivial steady states only occur for 
$c(t)\rightarrow  n(t)^2$ and one recovers the single-site
kinetic equation (rescaling time by a factor $(1-d)$)
\begin{equation} 
\dot{n} = (1-p)\, n^2 \left( 1 - n \right) -2p\, n^2 \label{MF}
\end{equation} 
which expresses that particles may only be created on empty sites.
The time-dependence of $n(t)$ for $t$ large is
\begin{equation} 
n(t) \simeq \left\{ 
\begin{array}{ll} 
n_{\infty} + \alpha \exp(-t/\tau) & \;; \mbox{\rm ~if $p<p_{c,{\sc mf}}(1)$ }\\
\sqrt{3/4} \cdot t^{-1/2}         & \;; \mbox{\rm ~if $p=p_{c,{\sc mf}}(1)$ }\\
1/(3p-1) \cdot t^{-1}             & \;; \mbox{\rm ~if $p>p_{c,{\sc mf}}(1)$ }
\end{array} \right.
\label{solMF}
\end{equation} 
where $n_{\infty}=(1-3p)/(1-p)$, $\tau=(1-p)/(1-3p)^2$, 
$p_{c,{\sc mf}}(1)=1/3$ and $\alpha$ is a constant
which depends on the initial conditions. At small $p$, where the 
creation process (\ref{2Ato3A}) dominates, the system is in the {\em active} 
phase with a non-vanishing particle density in the steady state. 
On the other hand, for $p$ large, the pair annihilation process (\ref{2Ato0}) 
dominates, the steady state particle density is zero and the system is in 
the {\em inactive} phase. 
In the entire inactive phase, and not only at the critical point,
the approach towards the steady state is algebraic, rather than exponential
as found in most systems. 

The same kind of result also holds in the pair approximation with $d\ne 1$. 
In the steady state, we have $c(\infty)=n(\infty)(1-3p)/(1-p)$. 
The particle density vanishes along the curve 
\begin{equation} 
p_{c,{\sc mf}}(d) = \left\{ \begin{array}{cl} 
\frac{1}{5} (1+3d)/(1-d) & ~;~ 0\leq d < 1/7  \\  
\frac{1}{3} & ~;~1/7< d\leq 1 
\end{array} \right. 
\end{equation} 
Close to the criticality, the particle density
$n(\infty) \sim p_{c,{\sc mf}}(d)-p$, which is the same found 
from $n_{\infty}$ in the $d\rightarrow 1$ limit above. However, 
the pair density
\begin{equation}  
c({\infty}) \sim \left\{ \begin{array}{cl} 
p_{c,{\sc mf}}(d)-p & ~;~ d< 1/7 \\
(p_{c,{\sc mf}}(d)-p)^2 & ~;~ d> 1/7 \end{array} \right. 
\end{equation}
There should thus be different universality
classes on both sides of the meeting (``tricritical'') 
point $p_{\rm t}=1/3$, $d_{\rm t}=1/7$, where   
$n(\infty)\sim (p_{\rm t}-p)^2$ and $c(\infty)\sim (p_{\rm t}-p)^3$ 
imply modified critical exponents. 

The leading long-time behaviour along the critical line $p=p_c(d)$ is
as follows
\begin{equation}
\left\{ \begin{array}{llll}
n(t) \sim t^{-1/2} &,& c(t) \sim t^{-1} & \mbox{\rm ~; if $d>d_{\rm t}$} \\
n(t) \sim t^{-1} &,& c(t) \sim t^{-3/2} & \mbox{\rm ~; if $d=d_{\rm t}=1/7$} \\
n(t) \sim t^{-1}   &,& c(t) \sim t^{-1} & \mbox{\rm ~; if $d<d_{\rm t}$} 
\end{array} \right. 
\end{equation}
while $n(t)\sim t^{-1}$ and $c(t)\sim t^{-2}$ in the entire inactive phase
$p>p_c(d)$. In the active phase, the steady state is approached exponentially.

Although generally believed to be qualitatively correct in sufficiently
high dimensions \cite{Priv96,Marr99,Hinr00}, 
kinetic equations cannot provide correct values of the exponents 
(and often not even the order of the transition) 
in low dimensions. In one dimension, the exact decay in the 
inactive phase is $n(t) \sim t^{-1/2}$ \cite{Howa97}, where the exponent 
$1/2$ is that of the process $2A \to \emptyset$ model \cite{Peli86}, which 
is recovered taking the limit $p \to 1$ in the rates (\ref{2Ato3A}) and 
(\ref{2Ato0}). However, the pair approximation suggests the presence of 
distinct universality classes for $d$ large and $d$ small. In addition, 
the exponent of the time-dependence of $n(t)$ for $d$ large equals $1/2$ as 
found exactly for one-dimensional diffusion. This suggests that the upper 
critical dimension $d^*$ for that transition should be unity, see \cite{Lee94},
but the value of $d^*$ in the PCPD is not yet known. To what extent are
the predictions of the pair approximation, in particular the existence of
several distinct universality classes along the transition line, borne out ?

An active state with a finite density of particles can be maintained only
in the limit $L \to \infty$. On a finite lattice any configuration of 
particles will decay towards an absorbing configuration in a finite time.
There are {\em two} possible absorbing configurations which the system 
cannot leave: (i) the empty lattice and (ii) a lattice occupied by one 
single diffusing particle. If we take $d=0$, we recover the pair contact 
process (PCP) introduced by Jensen \cite{Jens93}. In that case, any 
particle configuration without nearest-neighbour pairs is absorbing. The 
number of absorbing states grows exponentially with the chain length $L$ and
it is given by the Fibonacci number (see appendix A).
The steady-state properties of the $d=0$ phase transition between the active 
and inactive phases are described by the directed percolation (DP) universality
class \cite{Jens93}. However, the dynamical properties of this transition
are more subtle and still under active investigation \cite{Mend94,Dick99}. 
As we shall see, the presence of single-particle diffusion 
changes the universality class of the transition between the active and 
inactive phases: for any finite values of $d$ it does not fall in the DP 
universality class anymore.

Long ago, Janssen and Grassberger \cite{Jans81} conjectured that a model with a 
continuous phase transition from a fluctuating active phase into a phase with 
a single absorbing state and without additional symmetries is in the DP 
universality class. While it is widely believed that in the presence of local
symmetries in the reaction rates there should be a different universality 
class (normally the PC
if the particle number is locally conserved modulo 2), 
Park and Park \cite{Park95} provided a counterexample which shows
that even in the presence of local 
symmetries the steady-state transition can be in the DP 
universality class. This already indicates that the universality 
classification might be more subtle than previously thought. 
The exponents we want to compare with are (see \cite{Priv96,Marr99}) 
\begin{eqnarray} 
\mbox{\rm DP:}  \;\; &~& \theta=\nu_{\|}/\nu_{\perp}\simeq 1.5806\ldots\; ,\;\; 
\beta/\nu_{\perp} \simeq 0.2520\ldots \nonumber \\
\mbox{\rm PC:}
\;\; &~& \theta=\nu_{\|}/\nu_{\perp}\simeq 1.749\ldots\;\;\; 
,\;\; 
\beta/\nu_{\perp} \simeq 0.499 \ldots \label{eq:Expvals}
\end{eqnarray} 
and are defined as usual from the density $n(p) \sim (p-p_c)^{\beta}$
and the spatial and temporal correlation lengths $\xi_{\perp,\|} \sim
(p-p_c)^{-\nu_{\perp,\|}}$, see also section IV. 

The stochastic time evolution of the system is determined by the master 
equation, cast into the form
\begin{equation} 
\frac{\partial | P(t) \rangle}{\partial t} = - H | P(t) \rangle
\end{equation} 
where $| P(t) \rangle$ a state vector and $H$ is referred to as
``quantum'' Hamiltonian. For a chain with $L$ sites, 
$H$ is a stochastic $2^L \times 2^L$ matrix with elements
\begin{displaymath}
\langle \sigma |H |\tau \rangle = - w(\tau\to\sigma) \;\; , \;\;
\langle \sigma |H |\sigma\rangle = \sum_{\tau\neq\sigma} w(\sigma\to\tau)
\end{displaymath}
where $|\sigma\rangle$, $|\tau\rangle$ are the state vectors of the particle
configurations $\sigma,\tau$ and $w$ are the transition rates. 

Since $H$ is non-hermitian, it has distinct left and right eigenvectors.
We will use the notation $| 0_r \rangle$, $| 1_r \rangle$, \ldots, 
$| n_r \rangle$ ($\langle 0_l |$, $\langle 1_l |$, \ldots, $\langle 
n_l| $) for the right (left) eigenvectors of $H$ corresponding to energy 
levels $E_0$, $E_1$, \ldots, $E_n$ ordered according to $\Re E_0 \leq 
\Re E_1 \leq \ldots \Re E_n$, where $\Re$ denotes the real part.
One has $\langle n_l |m_r \rangle = 0$ if $E_n \neq E_m$ and we normalize 
the states in such a way that $\langle n_l |n_r \rangle = 1$.

Steady states are right eigenvectors of $H$ with zero eigenvalue. The two 
absorbing configurations mentioned above are steady states 
and given by
\begin{eqnarray} 
| 0_r \rangle &:=& | \emptyset \emptyset \emptyset 
\ldots \emptyset \rangle
\label{def0r} \\
| 1_r \rangle &:=&
| A \emptyset \emptyset \ldots \emptyset \rangle +
| \emptyset A \emptyset \ldots \emptyset \rangle + \ldots +
| \emptyset \emptyset \emptyset \ldots A \rangle 
\label{def1r}
\end{eqnarray}   
Thus, $H$ has two zero eigenvalues $E_0 = E_1 = 0$, while $\Gamma := \Re E_2$
is the inverse relaxation time towards the steady state.  
Furthermore, since $H$ is stochastic,  
\begin{equation} 
\langle 0_l | := \sum_\sigma \langle  \sigma |
\label{def0l}
\end{equation} 
is a left eigenvector of $H$ with zero eigenvalue. 

The calculation of the eigenvalues and eigenvectors of the stochastic 
Hamiltonian $H$, from which we will derive the critical properties of 
the model, is performed by the {\sc dmrg}\  algorithm \cite{Whit92,DMRG99} 
adapted to non-hermitian matrices, as described in detail in \cite{Carl99}.
In section III we present a further improvement 
to deal with systems with degenerate ground states.        

\section{Shift of the trivial ground state}
\label{sec:shf}

A severe numerical problem is generated by the fact that the physically
relevant eigenstate, the first excited state, is only the third state in the
spectrum due to the double degeneracy of the ground state. Asymmetric
diagonalization algorithms for large sparse matrices are much less robust
than their counterparts for symmetric matrices, making a precise determination 
of eigenvalues not at the extreme ends of the spectrum much more
demanding. In particular, the precision of the eigenstates suffers. This in
turn leads to a lower quality truncation of the state space in the {\sc dmrg}\ 
algorithm and subsequently to further deterioration of the results for longer
chains. In the present case, the {\sc dmrg}\  becomes unstable for rather short
chain lengths ($L \approx 20$). 

We alleviate this problem successfully by projecting out one of the two ground
states, for which both the left and right eigenstate are known,
as given in eqs. (\ref{def0l},\ref{def0r}). All {\em non-degenerate} 
eigenstates of the asymmetric Hamiltonian $H$ obey a biorthogonality condition.
This leaves us with two options to
eliminate one ground state: (i) Each diagonalization algorithm
for large matrices (Arnoldi or Lancz\'os) generates a sequence of
(bi)orthogonal trial vectors,
such that the requested eigenstate can be written as a linear combination
of those. During the generation, one may enforce orthogonality of all these
trial vectors with respect to the ground state. This option does not exist if 
one uses a black box routine. 
(ii) One may directly modify the Hamiltonian shifting the ground
state to an arbitrarily high energy, i.e. working with 
\begin{equation} 
H^\prime (\Delta) :=  H + \Delta | 0_r \rangle \langle 0_l |
\end{equation} 
where $\Delta$ is a positive number larger than the energy gap
of $H$. This new Hamiltonian is not stochastic, however it has the
same spectrum as $H$ apart from one of the ground states which is 
shifted to an energy level $\Delta$. The gap can now be obtained
from the first excited state of $H^\prime (\Delta)$.
This procedure can also be implemented for a stochastic Hamiltonian
with a single ground state. The gap is then obtained from the ground
state energy of $H^\prime (\Delta)$. This option can always be used
independently of the diagonalization method employed, see \cite{Mend99}
in relation with the power method.

To implement both options, one has to write down the left and right ground
states in the transformed block bases generated by the {\sc dmrg}\ . 

Let us denote for a block of length $L$ by $|\emptyset_{L}\rangle$ 
the state with no
particles and by $\langle\tau_{L}|\equiv\sum_{\sigma_{L}}\langle
\sigma_{L}|$
the sum over all states of a block (i.e.\ in the complete, unreduced Hilbert
space of the block). In the reduced block basis $\{ |m_{L}\rangle \}$ 
produced by the {\sc dmrg}\  we have
approximately $|\emptyset_{L}\rangle = \sum_{m_{L}}\langle
m_{L}|\emptyset_{L}\rangle |m_{L}\rangle$ and  
$\langle\tau_{L}| = \sum_{m_{L}}\langle
\tau_{L}|m_{L}\rangle \langle m_{L}|$. 
The right and left ground eigenstates
then read
\begin{eqnarray}
 & & |0_{r}\rangle_{2L+2} = |\emptyset_{L}\rangle \otimes |\emptyset\rangle 
 \otimes |\emptyset\rangle  \otimes |\emptyset_{L}\rangle = \nonumber \\ 
 & & \sum_{m_{L},m_{L}'} 
\langle m_{L}|\emptyset_{L}\rangle 
\langle m_{L}'|\emptyset_{L}\rangle 
(|m_{L}\rangle \otimes |\emptyset\rangle \otimes |\emptyset\rangle \otimes
|m_{L}'\rangle) , \label{eq:trafo1} \\
& & \langle 0_{l}|_{2L+2} = \sum_{s,s'}
\langle \tau_{L}| \otimes \langle s | 
 \otimes \langle s' | \otimes \langle \tau_{L}| = \nonumber \\ 
 & & \sum_{m_{L},m_{L}',s,s'} 
\langle \tau_{L}|m_{L}\rangle 
\langle \tau_{L}|m_{L}'\rangle 
(\langle m_{L}| \otimes \langle s| \otimes \langle s'| \otimes
\langle m_{L}'|) , \label{eq:trafo2}
\end{eqnarray}
where $s$ and $s'$ run over single site states. 

At the beginning of the {\sc dmrg}\  application, the matrix elements 
$\langle m_{L}|\emptyset_{L}\rangle$ and 
$\langle \tau_{L}|m_{L}\rangle$ are trivially constructed
for a sufficiently small $L$ such that
the Hilbert space has less then $m$ states, the number of states kept.

Using $\langle\tau_{L+1}| =
\sum_{s}\langle\tau_{l}| \otimes \langle s|$ and $|\emptyset_{L+1}\rangle =
|\emptyset_{L}\rangle \otimes |\emptyset\rangle$ and inserting one in 
$\langle \tau_{L+1}| = \sum_{m_{L+1}}\langle
\tau_{L+1}|m_{L+1}\rangle \langle m_{L+1}|$ and 
$|\emptyset_{L+1}\rangle = \sum_{m_{L+1}}\langle
m_{L+1}|\emptyset_{L+1}\rangle |m_{L+1}\rangle$ one finds as recursive
relation
\begin{eqnarray}
\langle m_{L+1}|\emptyset_{L+1}\rangle &=& \sum_{m_{L}}
\langle m_{L+1}| m_{L}\emptyset \rangle \langle m_{L}|\emptyset_{L}\rangle , \\
\langle \tau_{L+1}|m_{L+1}\rangle &=& \sum_{m_{L},s}
\langle m_{L}s| m_{L+1}\rangle \langle \tau_{L}|m_{L}\rangle .
\end{eqnarray}
The matrix elements $\langle m_{L+1}| m_{L}s\rangle$ 
are from the incomplete basis transformation $|m_{L}\rangle \otimes |s\rangle
\rightarrow |m_{L+1}\rangle$.

Numerical implementation of both approaches reveals that the second approach
is numerically very stable, while the first one is not, at least not if
one carries out an unsophisticated Gram-Schmidt orthogonalisation. We suppose
that this is due to the fact that global orthogonality of the trial vectors is
numerically not exactly conserved by the diagonalization algorithms, 
while Gram-Schmidt assumes this when a new trial vector is added and made 
orthogonal to the ground state. 
To find $\Gamma = \Re E_2$, we have chosen as density matrix
\begin{equation} 
\rho = \frac{1}{4} \widehat{\mbox{\rm tr}} 
\left\{ | 0_r \rangle \langle 0_r | + | 0_l \rangle \langle 0_l | +
| 2_r \rangle \langle 2_r | + | 2_l \rangle \langle 2_l | \right\}
\end{equation} 
where $\widehat{\mbox{\rm tr}}$ denotes the partial trace on the 
states of the left or right side of the chain.
It is essential to target also the ground state $|0_r\rangle, \langle 0_l|$
projected out to maintain a good
description of the ground state via (\ref{eq:trafo1}) and (\ref{eq:trafo2})
after some {\sc dmrg}\  steps. Otherwise, the Hamiltonian 
still contains a small 
perturbing contribution from the zero-energy ground state, which might 
destroy the stability of the procedure. 

\section{Numerical results}
\label{sec:pd}

\subsection{Analysis of the gap}

\subsubsection{Finite-size scaling method}

We are interested in the lowest energy gap
\begin{equation} 
\Gamma = \Gamma(p,d;L) = E_2
\label{defgap}
\end{equation} 
whose finite-size scaling behaviour contains the desired information 
about the critical properties of the PCPD model 
(\ref{2Ato3A},\ref{2Ato0},\ref{A0to0A}). 
While in principle the eigenvalues of the (non-symmetric) matrix $H$ 
may have a non-zero imaginary part, we always found that $E_2$ is real.

Generically, we expect the following finite-size behaviour of $\Gamma$
\begin{equation} 
\Gamma \sim \left\{ 
\begin{array}{ll}
\exp(-L/\xi_{\perp}) & \;; \mbox{\rm ~if $p<p_c(d)$} \\
L^{-\theta}          & \;; \mbox{\rm ~if $p=p_c(d)$} \\
L^{-2}               & \;; \mbox{\rm ~if $p>p_c(d)$}
\end{array} \right. \label{eq:GapSkal}
\end{equation}  
with $\theta = \nu_\|/\nu_\perp$ 
and $\xi_{\perp}$ the spatial correlation length. At a continuous
transition this diverges as $\xi_{\perp} \sim |p-p_c(d)|^{-\nu_{\perp}}$,
while in the time direction $\xi_\| \sim |p-p_c(d)|^{-\nu_{\|}}$.

The first line in (\ref{eq:GapSkal}) indicates that on a finite lattice, 
the relaxation towards the absorbing configurations is very slow. Indeed, 
for $L\rightarrow \infty$ the relaxation time $\tau=\Gamma^{-1}$ becomes 
infinite and a state with a finite density of particles becomes a steady state 
of the system. {From} the third line in (\ref{eq:GapSkal}) we see that 
in the entire inactive phase the scaling behaviour of the energy gap is 
algebraic, with an exponent equal to $2$, as for the pair-annihilation model 
$2A \to \emptyset$.
Therefore, in addition of the obvious double degeneracy of the ground state,
we find that our model is gapless in the large-system limit,
$\lim_{L\rightarrow \infty} \Gamma(p,d;L) =0$, for {\em all} values of 
$p$ and $d$ (in most systems the gap is finite in the inactive 
phase). 

The different phases and the critical point can be best identified from
the analysis of the quantity
\begin{equation} 
Y_L (p,d) := \frac{\ln \left[\Gamma(p,d;L+1)/\Gamma(p,d;L-1)\right]}
{\ln \left[(L+1)/ (L-1)\right]}
\label{defYL}
\end{equation} 
While usually, the critical point is found by looking for intersections
of two curves $Y_{L}$ and $Y_{L'}$ for two different lattice sizes $L,L'$,
it turns out that because of the scaling (\ref{eq:GapSkal}), {\em the critical
point is found from the maximum of $Y_L$ as a function of $p$ for fixed 
$d$ and $L$}. 
{From} (\ref{eq:GapSkal}) one then has for the large-$L$ behaviour of $Y_L(p,d)$
\begin{equation}  \label{eq:Ymaxval}
Y_L(p,d) \simeq \left\{ 
\begin{array}{ll}
-L/\xi_{\perp}  & \;; \mbox{\rm ~if $p<p_c(d)$} \\
-\theta         & \;; \mbox{\rm ~if $p=p_c(d)$} \\
-2              & \;; \mbox{\rm ~if $p>p_c(d)$}
\end{array} \right.
\end{equation} 
Therefore, the location of the maximum value of $Y_L(p,d)$ yields 
a sequence of estimates $p_c(d;L)$ and the critical point can be found 
from extrapolating these sequences to $L \to \infty$.
We point out, however, that the value of $Y_L$ at its maximum {\em cannot}
be used to infer the value of the exponent $\theta$. Only after $p_c(d)$ is
determined for the $L\to\infty$ lattice, we can use (\ref{eq:Ymaxval}) to
find $\theta$. The technicalities of the method are discussed in 
appendix B \cite{danke}. 

\subsubsection{Inactive phase}

\begin{figure}[bt]
\centerline{\psfig{file=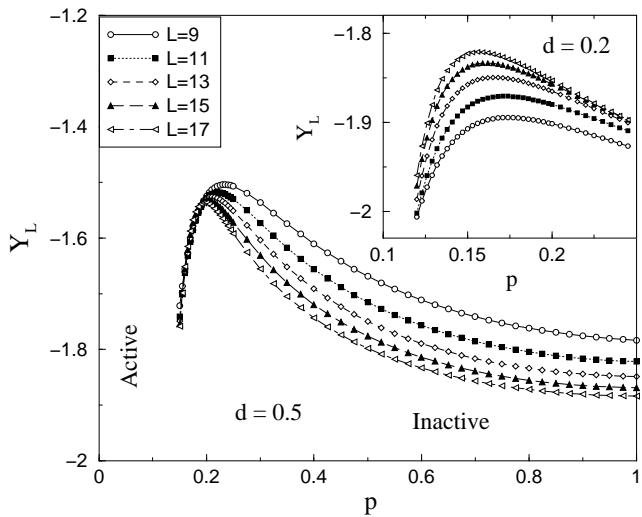,height=7cm}}
\vskip 0.2truecm
\caption{Plot of $Y_L(p,d)$ as function of $p$ for various lattice sizes 
and $d = 0.5$. The inset shows the same quantity for $d=0.2$.}
\label{YLd0.5}
\end{figure}

Figure \ref{YLd0.5} shows a plot of $Y_L(p,d)$ as function of the parameter
$p$ and for $d=0.5$ for $L=9$, $11$, $13$, $15$ and $17$ 
(in the inset we show $Y_L$ for $d=0.2$). 
Calculations were performed up to $L=30$, but for the largest sizes only in 
the vicinity of the critical point. 

In the active phase (small $p$) the quantity $Y_L(p,d)$ scales linearly with 
$L$ as predicted in (\ref{eq:Ymaxval}). For larger $p$ the reaction $2A \to 
\emptyset$ dominates and the system is in the inactive phase, for which we 
should have asymptotically $Y_L(p,d) \simeq -2$.
As the system size $L$ is increased, the curves approach the expected value 
$-2$ in the whole inactive phase. Table \ref{BSTinactive} shows the values 
of $-Y_L(p,d)$ calculated for three different diffusion constants and in the 
inactive phase. We extrapolated the finite-lattice data, using the {\sc bst} 
algorithm \cite{Buli64} and found excellent agreement with
the expected value of $\theta=2$, see eq.~(\ref{eq:Ymaxval}). 
The convergence is similar to the examples studied in \cite{Carl99},
although the raw data can be quite far from their $L\to\infty$ limit value.
This check also confirms that the numerical values for the energy gap,
as obtained from the {\sc dmrg}\  calculation, are very accurate.

\vbox{
\begin{table}[tb]
\caption{Finite-size estimates $-Y_L$ of the dynamical exponent $\theta$
and their $L\to\infty$ extrapolation, 
obtained by the {\sc bst}\  method, in the inactive phase. 
\label{BSTinactive}
}
\vskip 0.2truecm
\begin{tabular}{cccc}
$L$ & $d = 0.1$, $p = 0.6$ & $d = 0.5$, $p = 0.6$ & $d = 0.8$, $p = 0.5$  
\\ \hline
   9  & 1.979223431 & 1.7110633 & 1.226133000 \\
  11  & 1.986196898 & 1.7567773 & 1.267952170 \\
  13  & 1.990160149 & 1.7893116 & 1.306670859 \\
  15  & 1.992629239 & 1.8139503 & 1.342357293 \\
  17  & 1.994271913 & 1.8333674 & 1.375169517 \\
  19  & 1.995420176 & 1.8491053 & 1.405319671 \\
  21  & 1.996254444 & 1.8621348 & 1.433037552 \\
  23  & 1.996879687 & 1.8731052 & 1.458550029 \\
  25  & 1.997360512 & 1.8905590 & 1.482070754 \\
  27  & 1.995400349 & 1.8976147 & 1.503795737 \\
  29  &             &           & 1.523902146 \\ \hline
$\infty$&2.0000(1)  & 2.000(1)  & 1.99(2) \\ 
\end{tabular} \end{table}
}

\subsubsection{Active-inactive transition line}

Table~\ref{tab:param} collects our results for the critical point
$p_c(d)$ and the exponents $\theta$ and $\beta/\nu_{\perp}$, for several
values of $d$. 

\vbox{
\begin{table}
\caption{Critical parameters $p_c(d)$, $\theta$ and $\beta/\nu_{\perp}$ 
along the active-inactive transition line for several values of $d$. 
\label{tab:param}}
\begin{tabular}{c|llllll}
$d$      & 0.10     & 0.15     & 0.20     & 0.35     & 0.50     & 0.80  \\\hline
$p_c$    & 0.111(2) & 0.116(2) & 0.121(3) & 0.138(1) & 0.154(1) & 0.205(3) \\
$\theta$ & 1.87(3)  & 1.84(3)  & 1.83(3)  & 1.72(3)  & 1.70(3)  & 1.60(5)  \\
$\beta/\nu_\perp$ 
         & 0.50(3)  & 0.49(3)  & 0.49(3)  & 0.47(3)  & 0.48(3)  & 0.51(3)
\end{tabular}
\end{table}
}

First, estimates for $p_c(d)$ in the $L\to\infty$ limit 
were obtained from a linear fit in $1/L$ of the finite-size data $p_c(L)$, 
i.e. the abscissas of the maxima of $Y_L (p,d)$. 
For comparison, we recall that $p_c(0)=0.077090(5)$ \cite{Dick98}. 

\begin{figure}[b]
\centerline{\psfig{file=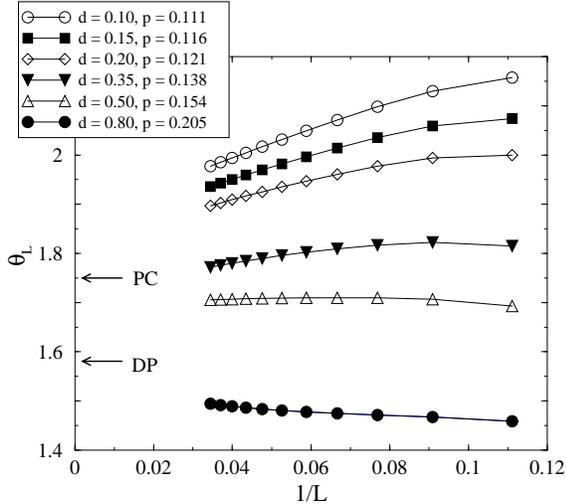,height=7cm}}
\vskip 0.2truecm
\caption{Plot of $\theta_L$ as a function of $1/L$ for various values of $d$.
The two horizontal arrows point to the values of $\theta$ expected for the
DP ($\theta \approx 1.58$) and PC ($\theta \approx 1.75$) universality classes.
}
\label{thetaL}
\end{figure}

Next, we estimate the value of the exponent $\theta$ from the extrapolation
of $\theta_L = -Y_L(p_c(d), d)$ to the thermodynamic limit $L \rightarrow  
\infty$. Figure \ref{thetaL} shows a plot of $\theta_L$ as a function of 
$1/L$ for various values of $d$. 
We notice a different finite-size scaling behaviour of $\theta_L$ for $d \leq
0.20$ and for $d=0.50$. In the former case $\theta_L$ varies quite strongly 
with $L$, starting from a value above $2$ and going towards values below $2$.
For $d \approx 0.5$, $\theta_L$ shows little $L$-dependence. 
The extrapolated values of $\theta$ (coming from a cubic fit in $1/L$) 
are shown in Table \ref{tab:param}.
Error bars are due to the uncertainty in the determination of the critical
point location. 
The final estimates for $\theta$ appear to vary with $d$.
It is not yet clear whether this variation is real or the consequence of an
unresolved finite-size correction term. However, 
the results are clearly inconsistent with a transition of DP type, for which 
$\theta \approx 1.58$. 

\subsection{Steady state particle density}

For a finite lattice and for all values of $p$ and $d$ the stationary
states, given in (\ref{def0r}) and (\ref{def1r}), contain either no 
or just one particle. Therefore the steady-state particle-density vanishes 
in the large $L$ limit. 

To obtain a steady state with a finite density of particles we added 
a particle creation process at the two boundary sites \cite{Carl99}
\begin{equation}
\emptyset \to A \,\,\,\, {\rm with \,\,\,rate}\,\, p',
\label{boundreac}
\end{equation}
We are interested in the (steady state) density profile, 
for a chain of length $L$ 
\begin{equation}
n_L (l) = \langle 0_l | \hat{n}(l) | 0_r \rangle
\end{equation} 
where $l=1,2,\ldots L$ labels the position along the chain,
$\hat{n}(l)$ is the particle number operator at site $l$ and 
$| 0_r \rangle$ and $\langle 0_l |$ are the left and right ground 
states of the non-symmetric operator $H$, with the terms coming 
from (\ref{boundreac}) added. 

The boundary reaction term removes the ground state 
degeneracy of $H$ encountered earlier. Therefore, {\sc dmrg}\  calculations 
are easier to perform when $p'\neq 0$. In this case it is not necessary to 
follow the shift strategy introduced in Sec. \ref{sec:shf}. 
In fact the {\sc dmrg}\  calculations are stable up to chains of lengths 
$L \approx 50-60$, i.e. almost of the double size with respect of the lengths 
we could reach in the study of the energy gap.

In general, the time-dependent particle density $n=n(t,p;l,L)$ at site $l$
for a lattice of size $L$ and in the vicinity of the critical point,
should satisfy the scaling form
\begin{eqnarray} 
n(t,p;l,L) &=& t^{-\beta/\nu_{\|}} 
F\left({t}/{\xi_{\|}},{L}/{\xi_{\perp}},{l}/{L} \right) \nonumber\\
&\sim& \xi_{\perp}^{-\beta/\nu_{\perp}} 
G\left({t}/{\xi_{\perp}^{\theta}},{L}/{\xi_{\perp}},{l}/{L}\right)
\label{eq:DensSkal}
\end{eqnarray} 
where for simplicity the dependence on $d$ and $p'$ is suppressed and 
$F$ and $G$ are scaling functions. The exponents have their usual 
meaning \cite{Marr99}. 
In particular, the steady-state density profile, at the critical point,
should satisfy
\begin{equation}
n_L (l) := n(\infty,p_c;l,L) =  L^{-\beta/\nu_\perp} f \left( l/L \right)
\label{scaldens}
\end{equation}
with some scaling function $f(z)$. 

\begin{figure}[b]
\centerline{\psfig{file=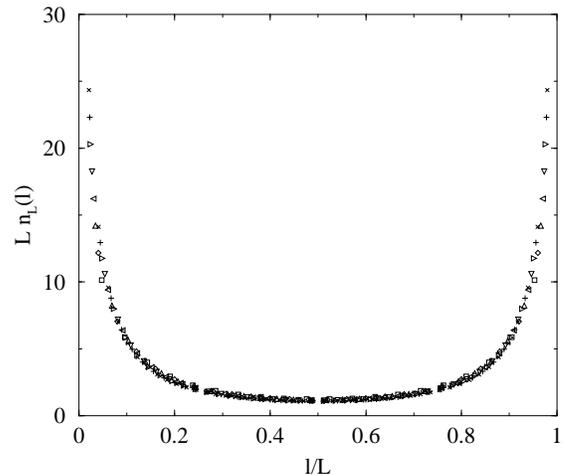,height=6.5cm}}
\vskip 0.2truecm
\caption{Scaled particle density $L n_L (l)$ as function of the scaled
variable $l/L$ in the inactive phase for $d=0.5$, $p = 0.8$ and
injection rate $p' = 0.3$. Data for different system sizes 
($L = 20$, $24$, $28$, \ldots $48$) collapse nicely onto a single curve.}
\label{density}
\end{figure}

First, we consider the inactive phase. For our model, the average 
particle density decays algebraically for large times as ${n}(t) 
\sim t^{-1/2}$. From (\ref{eq:DensSkal}), we identify $\beta/\nu_\| = 1/2$. 
Thus $\beta/\nu_\perp =\theta\beta/\nu_\| = 1$ since the anisotropy 
exponent $\theta = \nu_\|/\nu_\perp = 2$ in the inactive phase,
see table~\ref{BSTinactive}. 

The scaling (\ref{scaldens}) is confirmed in Fig. \ref{density}, obtained
for chains up to $L=48$, which shows the scaled particle density $L n_L(l)$
as a function of $l/L$. 
The ratio $\beta/\nu_{\perp}=1$ for the whole inactive phase has also 
been confirmed with good accuracy from {\sc bst}\  extrapolations.

{From} now on we focus on the transition between the active and inactive 
phases. We concentrate on the {\em central} particle density
\begin{equation} 
n(L) := n_L (L/2)
\end{equation}  
For this quantity we expect that for $L$ large, $ n(L) \simeq n_0 + 
O(e^{-L/\xi_{\perp}}) > 0$ in the active phase where at the critical point
\begin{equation}
n(L) \sim L ^{-\beta/\nu_\perp}
\label{scalnL}
\end{equation}
while, as shown above, $n(L) \sim L^{-1}$ in the entire inactive phase. 

\begin{figure}[b]
\centerline{\psfig{file=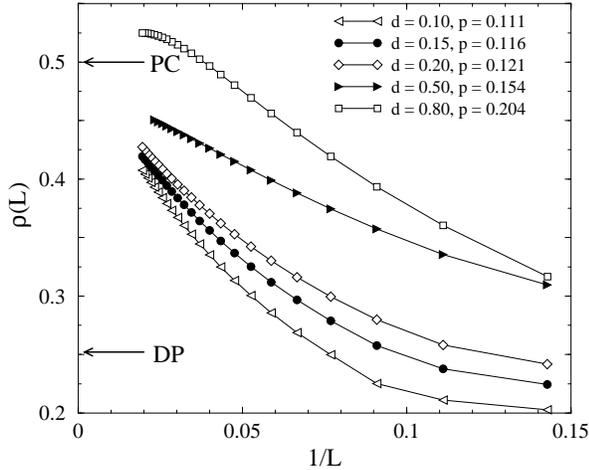,height=6.5cm}}
\vskip 0.2truecm
\caption{
The effective exponent $\rho(L)$ as function of $1/L$ for various
values of $d$ and up to $L=51$. The horizontal arrows point to the values 
expected for DP and for PC universality classes.
}
\label{rhoL}
\end{figure}

Again, in analogy with (\ref{defYL}) we form the logarithmic derivative
\begin{equation}
\rho(L) = - \frac{\ln [n(L+1)/n(L-1)] }{\ln [(L+1)/(L-1)]}
\end{equation}
According to eq. (\ref{scalnL}), $\rho(L)$ is expected to converge to the
ratio $\beta/\nu_\perp$ for $L \to \infty$ {\em at} the critical point 
$p=p_c(d)$ as listed in Table~\ref{tab:param}.
A plot of $\rho (L)$ for various values of the diffusion constant is shown in
Fig. \ref{rhoL}. 
We notice that at small $L$ and $d$ the effective exponent $\rho(L)$ starts
from values close to DP, while it clearly deviates from it for increasing $L$.
A cubic fit in $1/L$ yields the extrapolated exponent ratio $\beta/\nu_\perp$
as a function of $d$, see Table~\ref{tab:param}. This ratio appears to be 
rather constant with $d$.
As in the case of the calculation of $\theta$, we notice that the exponents
are clearly inconsistent with a DP transition. However, the extrapolated
values are close to the exponent value expected for a transition in the PC 
class.

\section{Discussion}
\label{sec:dis}

\begin{figure}[b]
\centerline{\psfig{file=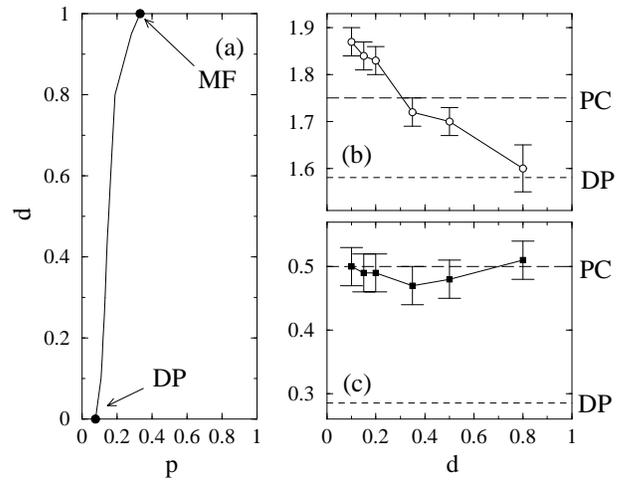,height=6.5cm}}
\vskip 0.2truecm
\caption{(a) Steady-state phase diagram of the model $2A \to 3A$, $2A \to 
\emptyset$ as determined by {\sc dmrg} techniques. The continuous line 
separates the active from the inactive phase. Extrapolated estimates of the 
exponent $\theta$ are shown in (b) and of the exponent $\beta/\nu_{\perp}$ 
in (c) as a function of $d$.}
\label{phasediagram}
\end{figure}

We collect the results of this study of the Pair Contact Process with Diffusion.
In Fig. \ref{phasediagram}(a) we show the steady-state 
phase diagram of the PCPD model. 
For small enough $p$, there is an active phase with a non-vanishing
steady-state particle density and which goes over into an inactive state at
some critical point $p_c(d)$.  
The critical line separating the active from the inactive
phases terminates for $d \to 0$ at 
the DP point $p_c(0) = 0.077090(5)$ \cite{Dick98}.
For $d \to 1$ the critical line terminates at the MF point $p = 1/3$,
as predicted from the mean field equations. This should have been expected, 
since it is well-known that when diffusion dominates over all other reactions
the critical behaviour becomes of mean field type, even in one dimension 
\cite{Marr99,Hinr00,Bethe-Ansatz}.

In the PCPD, the entire inactive phase is critical and is expected to be 
in the universality class of diffusion-annihilation \cite{Howa97}. 
We have found the exponents $\theta=2$ and $\beta/\nu_{\perp}=1$,  
confirming this expectation. 

We have investigated the properties of the transition line between the active
and inactive phases. Fig. \ref{phasediagram}(b) and (c) show the numerical 
estimates of the exponents $\theta$ and $\beta/\nu_\perp$, respectively, 
see also Table~\ref{tab:param}. For comparison, the values 
(\ref{eq:Expvals}) of these exponents for the DP and PC universality 
classes are shown as horizontal lines. Clearly, our results are incompatible 
with a transition in the DP class. That means that single particle diffusion is 
a relevant perturbation of the pair contact process. 
While $\beta/\nu_\perp$ is quite 
constant and consistent with the PC value within error bars, $\theta$ seems 
to vary continuously with $d$. 

There is no clear evidence for two distinct universality classes along the
critical line, as predicted by pair mean field theory (see section II), but
our information on the transition line for $d$ close to unity remains 
incomplete (see also below).  
At present, neither of the two possibilities can be ruled out. 
It is an open question whether the $2D$ PCPD model phase diagram
will be in qualitative agreement with pair mean field theory.

Turning to the apparent variation of $\theta$ with $d$, note that $\theta$
decreases systematically with $d$, starting from $\theta\approx 2$ in the
$d\to 0$ limit. To explain this, consider the case when $d$ is small and
the particle density is low, i.e. the system is close to criticality. 
Then the particles will diffuse independently until they meet and react. 
The lower the diffusion rate $d$, the larger the time interval $t_{\rm diff}$
for which free diffusion can be observed. Associated to this is a length
scale $R_{\rm diff}$, viz. $t_{\rm diff}\sim R_{\rm diff}^2$. For times
short compared to $t_{\rm diff}$, the system is governed by diffusion only. 
Therefore, for system sizes $L\ll R_{\rm diff}$, the critical exponents
will be effectively those of free diffusion and 
$\theta\simeq \theta_{\rm diff}=2$. The true value for $\theta$ will be seen
only if $L\gg R_{\rm diff}$. These two regimes are separated by a crossover
region and it appears plausible that the apparent variation of our estimates of
$\theta$ with $d$ might be the consequence of such a crossover. 
However, this crossover phenomenon does not show up in the calculation of 
$\beta/\nu_\perp$. That is to be expected
since by injecting particles at the boundaries we induce a {\em finite} 
density of particles also at the critical point and $t_{\rm diff}$ will 
no longer increase beyond any bounds as $d\to 0$ and the above heuristic 
argument no longer applies. 

While the exponent $\beta/\nu_\perp$ is not far from the one of the PC 
class it is difficult to extract reliable information form $\theta$.
We have no reason to believe that exponents should be continuously varying
as function of the diffusion constant. A closer inspection of the finite-size 
behaviour (see Fig. \ref{thetaL}) reveals that the $L\to\infty$ 
asymptotic value $\theta$ 
is approached from above for $d \leq 0.35$, while it is approached from
below for $d \geq 0.5$. As finite-size effects are seen to be weak in
the interval $0.35 \leq d \leq 0.5$, we presume that the most reliable
estimates for $\theta$ might be those obtained in this range of diffusion
constants. Therefore most likely $\theta \approx 1.7$, which is not far
from the PC value either.
We also recall that the relaxation in the inactive phase is algebraic:
All known models \cite{Park95,Gras84,Meny94,Inui98,Hinr97} 
in the PC class are characterized by an algebraic relaxation
in the inactive phase with exponents $\theta=2$ and $\beta/\nu_{\perp}=1$, 
as for the PCPD model.
Therefore, it might be tempting to speculate that the active-inactive 
transition of the PCPD model were in the PC universality class. 

On the other side all known models in the PC universality class are 
characterized by some conservation laws either on the parity of the 
number of particles or by an exact symmetry between the absorbing states 
\cite{Hinr00}. Such local consevation laws are however lacking for the 
PCPD model, which suggests that the PCPD model should not be in 
the PC class.
It should be also stressed that the unambigous identification of a steady-state
universality class requires the determination of {\it four} independent 
exponents, see \cite{Marr99,Hinr00}, while our techniques provided values for 
only two. 

In summary, we have used the {\sc dmrg} method to find the steady-state 
phase diagram of the PCPD model. Single particle diffusion is a relevant
perturbation for the system since the model does not show a DP behaviour
as in the limit of vanishing diffusion constant.

Some exponent values along the active-inactive 
transition line are surprisingly close to those of the PC universality class.
However, it is conceivable that the near coincidence of the exponents with 
those of the PC class might be accidental. 
In that case the transition(s) would belong to a universality 
class distinct form both DP and PC. 
It is not yet clear whether in the $1D$ PCPD there are two 
distinct transitions, as suggested by pair mean field theory, 
or merely a single one. 
All in all, complementary studies would be needed to fully understand the 
remarkably subtle behaviour of this so simple-looking model. \\

\noindent{\bf Note added:} After this paper was submitted, 
Hinrichsen \cite{Hinr00a} performed a Monte
Carlo study of the PCPD for $d=0.1$. The time-dependent density
$n(t)\sim t^{-\delta}$ is characterized by the exponent 
$\delta=\beta/(\nu_{\perp}\theta)$. He finds $\delta=0.25(2)$ 
and $\theta=1.83(5)$. This agrees with our Table~\ref{tab:param}. 
After this paper was accepted, we also received
a paper by \'Odor \cite{Odor00} which studies the PCPD through Monte Carlo
simulations and the coherent anomaly method. 
In particular, for $0.05\leq d\leq 0.2$, \'Odor finds
$\delta\approx 0.27$ and for $d=0.5$ and $d=0.9$, $\delta\approx 0.2$. 
The first result compares well with our results from Table \ref{tab:param},
which gives $\delta\approx 0.27$ 
and agrees with the result of Hinrichsen \cite{Hinr00a}. 
However, \'Odor also finds $\beta\approx 0.58$ for $0.05\leq d \leq 
0.2$, consistent with the upper bound $\beta <0.67$ reported in 
\cite{Hinr00a}. That is far away from the PC value $\beta_{\sc pc}
\simeq 0.93$ \cite{Marr99,Hinr97}.

\acknowledgements

It is a pleasure to thank P. Grassberger, H. Hinrichsen, M. Howard, 
J.F.F. Mendes, G. \'Odor, U. T\"auber and F. van Wijland for many stimulating 
discussions and J.F.F. Mendes for participation in the 
earliest stages of this work. M.H. thanks the Department of Theoretical Physics
of the University of Oxford for warm hospitality and J.L. Cardy and Z. Racz for 
discussions, where early stages of this work were performed. We thank the 
Centre Charles Hermite in Nancy for providing substantial computational time. 
E.C. acknowledges partial financial support from a grant fo the Minist\`ere 
des Affaires Etrang\`eres No. 224679A. M.H. and U.S. were supported in part 
by the Procope programme. 
                                                     
\section*{Appendix A: Number of Absorbing Configurations in the PCP}

For $d=0$, the model reduces to the pair contact
process (PCP) \cite{Jens93}. In this case, all configurations of the type 
$| \ldots \emptyset A \emptyset \ldots \emptyset A \emptyset \ldots \rangle$, 
without nearest neighbour particles are absorbing and stationary 
states. We find the number $N(L)$ of absorbing states in the PCP for 
a chain of length $L$ with free (periodic) boundary conditions.

For {\em free} boundary conditions, one has the recursion
\begin{equation} 
N(L) = N(L-1) + N(L-2).
\label{recursion}
\end{equation}  
To see this, concentrate on the leftmost site. If it is occupied, its
neighbour must be empty for the state to be absorbing and there remains an 
open chain of $L-2$ sites to be considered. If it is empty, one considers the
open chain of the remaining $L-1$ sites. 
The initial conditions for the problem are $N(1)=2$ and $N(2)=3$.
Therefore $N(L) = F_{L+1}$ is the $(L+1)^{\rm th}$ Fibonacci number. 

This can also be seen from the generating function
\begin{equation} 
\tilde{N}(s) = \sum_{k=0}^{+ \infty} N(k) s^k
\label{series}
\end{equation} 
which satisfies because of (\ref{recursion})
\begin{equation} 
\tilde{N}(s) = \frac{N(0) (1-s) + N(1) s}{1 - s - s^2 }
 = \frac{1+s}{1 - s - s^2 }
\end{equation} 
For the inverse transformation one can use
\begin{equation} 
N(L) = \frac{1}{2 \pi {\rm i}} \oint {\rm d}s \,\, \tilde{N}(s) \,\, s^{-L-1}
\end{equation} 
where the integral is taken in a closed circle centered in the
origin of the complex $s$-plane and with radius smaller than the
radius of convergence of the series (\ref{series}), so as to
exclude contributions of other poles than that in $s = 0$.
The transformation $z = 1/s$ yields
\begin{equation} 
N(L) = \frac{1}{2\pi {\rm i}}\oint {\rm d}z\, \frac{1+z}{z^2-z-1}z^L 
= \frac{z_+^{L+2} - z_-^{L+2}}{z_+ - z_-}
\end{equation} 
where $z_\pm = (1 \pm \sqrt 5)/2$; $z_+$ is the golden mean.  

For {\em periodic} boundary conditions, the number $N_{\rm per}(L)$ of 
absorbing states for a chain of $L$ sites is 
\begin{equation} 
N_{\rm per} (L) = N(L-1) + N(L-3) = z_+^L + z_-^L
\end{equation} 
For $L$ large the asymptotic behaviour is $N(L) \sim N_{\rm per}(L) 
\sim z_+^L$. 

Another example of a kinetic model with exponentially many absorbing states
is an adsorption-desorption model of $k$-mers ($k\geq 3$) on a $1D$ lattice 
where the number of absorbing states is described by generalised Fibonacci
numbers \cite{Barm93}.   

\vspace{-2mm}
\section*{Appendix B: Finite-size-scaling in systems with a critical
phase}

We discuss the finite-size scaling of the lowest gap $\Gamma=\Gamma_L(p)$ and 
how to localize the critical point. For simplicity, we suppress the dependence 
on $d$ and write the scaling form
\begin{equation} 
\Gamma_L(p) = L^{-\theta} f\left( (p-p_c) L^{1/\nu_{\perp}} \right)
\end{equation} 
where $f$ is assumed to be continously differentiable. 
For the gap, one expects the behaviour
\begin{equation}  \label{eq:GammaSkal}
\Gamma_L(p) \sim \left\{ \begin{array}{ll}
e^{-\sigma L} & \;; \mbox{\rm ~if $p<p_c$} \\
L^{-2}        & \;; \mbox{\rm ~if $p>p_c$, case C} \\
\Gamma_{\infty} & \;; \mbox{\rm ~if $p>p_c$, case N}
\end{array} \right.
\end{equation} 
where $\sigma, \Gamma_{\infty}$ are constants independent of $L$. 
Here, case N refers to the ``normal'' case of non-critical phases 
on both sides of the transition at $p=p_c$ and case C alludes to 
the case of a critical phase on one side and we have already set 
the exponent equal to 2 in view eq.~(\ref{eq:GapSkal}) valid for 
our model. This implies for the scaling function $f(z)$
\begin{equation} 
f(z) \sim \left\{ \begin{array}{ll} 
\exp\left(-A |z|^{\nu_{\perp}}\right) 
& \;;\mbox{\rm ~$z\rightarrow -\infty$}\\
z^{(\theta-2)\nu_{\perp}} & \;; \mbox{\rm ~$z\rightarrow +\infty$, case C} \\
z^{\theta\nu_{\perp}} & \;; \mbox{\rm ~$z\rightarrow +\infty$, case N}
\end{array} \right.
\end{equation} 
where $A$ is a positive constant. Since $f(z)$ is positive, 
it follows that in the case C with $\theta<2$, 
$f(z)$ must have a maximum at some finite value $z_{\rm max}$. 
For the case N, however, $f(z)$ should increase monotonically with $z$. 

The estimator $Y_L$ as defined in (\ref{defYL}) then becomes
\begin{equation}  \label{eq:Ywert}
Y_L = -\theta + \frac{\ln\left[ f(z_+)/f(z_-) \right]}{\ln[(L+1)/(L-1)]}
\end{equation} 
where $z_{\pm}=(p-p_c)(L\pm1)^{1/\nu_{\perp}}$. Furthermore, writing
$g(z) = \ln f(z)$, a straightforward calculation gives
\begin{eqnarray} 
\lefteqn{ \lim \frac{{\rm d}Y_L}{{\rm d}p} = 
\frac{L^{1/\nu_{\perp}}}{\nu_{\perp}}
\left[ g'(z) - zg''(z) \right] } \\
& \simeq & \left\{ \begin{array}{ll} 
L^{1/\nu_{\perp}} A (2-\nu_{\perp}) (-z)^{\nu_{\perp}-1} & ~;~ z\to-\infty \\
L^{1/\nu_{\perp}} (\theta-2) z^{-1} & ~;~ z\to\infty
\end{array} \right. \nonumber
\end{eqnarray} 
in the finite-size scaling limit $p\rightarrow  p_c$ and $L\rightarrow \infty$ 
simultaneously such that $z=(p-p_c)L^{1/\nu_{\perp}}$ is kept fixed. 
For the case C with $\theta<2$ and $\nu_{\perp}<2$, there is some finite 
$z^*$ such that ${\rm d}Y_L/{\rm d}p|_{z=z^*} =0$. Then
\begin{equation} \label{gl:Y}
Y_L(z^*) = -\theta + \frac{1}{\nu_{\perp}} z^* g'(z^*)
\end{equation}
If we choose $z=z^*$, we have a sequence of values of $p_L$ 
converging towards $p_c$ according to $p_L\simeq p_c + z^* L^{-1/\nu_{\perp}}$.
On the other hand, {\em $p_L$ can be found by determining the maximum of $Y_L$ 
as a function of $p$}, since $\lim {\rm d}Y_L/{\rm d}p|_{z^*}=0$. 
This is the desired result. However, because of (\ref{gl:Y}) $Y_L(z^*)$ does
{\em not\/} readily yield an estimate for the exponent $\theta$, since there is 
no guarantee that $g'(z^*)$ should vanish, (or in other words that 
$z^*=z_{\rm max}$). The generalization to other observables
with scaling analogous to (\ref{eq:GammaSkal}) is immediate. 

For the case N, however, this technique does not apply, since in general
$f'(z)>0$ for all values of $z$.  

Finally, we recall that the leading finite-size correction terms determine
whether or not the curves $Y_L(p)$ and $Y_{L'}(p)$ will intersect. Consider the
extended scaling form $\Gamma_L(p) = L^{-\theta} f(z) [ 1+L^{-\omega} A(z)]$,
where $\omega>0$ is the leading correction exponent. If $\omega<2$, we find
\begin{equation} \label{gl:YK}
Y_L = -\theta+\frac{z g'(z)}{\nu_{\perp}} + L^{-\omega} A(z) \left(
-\omega + \frac{1}{\nu_{\perp}}\frac{z A'(z)}{A(z)} \right) 
\end{equation}
up to terms of order $O(L^{-2},L^{-1-\omega})$. 
Now, the curves $Y_L$ and $Y_{L'}$ intersect if there is some $z_{\rm int}$ such
that the scaling function of the leading correction term in (\ref{gl:YK}) 
vanishes. But that term depends {\em only} on the correction amplitude $A(z)$
and is independent of $f(z)$.

\end{multicols}

\vspace{-4mm}
\begin{references}
\vspace{-16mm}
\bibitem[*]{UMR} Unit\'e Mixte de Recherche CNRS No~7556
\bibitem{Priv96} V. Privman (Ed.) {\it Nonequilibrium Statistical Mechanics in
One Dimension}, Cambridge University Press (Cambridge 1996)
\bibitem{Marr99} J. Marro and R. Dickman, {\it Nonequilibrium Phase Transitions
in Lattice Models}, Cambridge University Press (Cambridge 1999)
\bibitem{Hinr00} H. Hinrichsen, Adv. Phys. {\bf 49}, 1 (2000)
\bibitem{Card96} J. Cardy and U. C. T\"auber, \prl {\bf 77}, 4780 
(1996) and J. Stat. Phys {\bf 90}, 1 (1998); 
D.C. Mattis and M.L. Glasser, Rev. Mod. Phys. {\bf 70}, 979 (1998)
\bibitem{Bethe-Ansatz} D. Kandel, E. Domany and B. Nienhuis, J. Phys. A:
Math. Gen. {\bf 23}, L755 (1990); 
F.C. Alcaraz, M. Droz, M. Henkel and V. Rittenberg, 
Ann. of Phys. {\bf 230}, 250 (1994); 
I. Peschel, U. Schultze and V. Rittenberg, Nucl. Phys. {\bf B430}, 633 (1995); 
P.-A. Bares and M. Mobilia, \prl {\bf 83}, 5214 (1999) and {\bf 85}, 893 (2000);
G.M. Sch\"utz, {\it Integrable Stochastic Many-Body Systems}, 
to appear in C. Domb and J.L. Lebowitz (Eds.), {\it Phase Transitions and 
Critical Phenomena}, Vol. 19, Academic Press (New York 2000).
\bibitem{Chop98} B. Chopard and M. Droz, {\it Cellular Automata Modelling
of Physical Systems}, Cambridge University Press (Cambridge 1998)
\bibitem{Henk90} M. Henkel and H. Herrmann, J. Phys. A: Math. Gen. {\bf 23},
3719 (1990); J.R.G. de Mendon\c{c}a, \pre {\bf 60}, 1329 (1999).
\bibitem{Mend99} J.R.G. de Mendon\c{c}a, J. Phys. A Math. Gen. {\bf 32}, 
L467 (1999).
\bibitem{Carl99} E. Carlon, M. Henkel and U. Schollw\"{o}ck, Eur. Phys.
J. {\bf B12}, 99 (1999).
\bibitem{Hooy99} J. Hooyberghs and C. Vanderzande, J. Phys. A: Math. Gen. 
{\bf 33}, 907 (2000). 
\bibitem{Howa97} M.J. Howard and U.C. T\"auber, J. Phys. A:
Math. Gen. {\bf 30}, 7721 (1997).
\bibitem{Jens93} I. Jensen, \prl {\bf 70}, 1465 (1993).
\bibitem{Dick98} R. Dickman and J. Kamphorst Leal da Silva, Phys. Rev. 
{\bf E58}, 4266 (1998).
\bibitem{Whit92} S.R. White, Phys.\ Rev.\ Lett.\ {\bf 69}, 2863 (1992);
Phys.\ Rev.\ {\bf B 48}, 10345 (1993).
\bibitem{DMRG99} I. Peschel, X. Wang, M. Kaulke and K. Hallberg (Eds.):
{\em Density Matrix Renormalization: A New Numerical Method in Physics}, 
Springer, (Heidelberg 1999).
\bibitem{Burs96} R.J. Bursill, T. Xiang and G.A. Gehring, J. Phys. Cond. Mat.
{\bf 8}, L583 (1996);
X. Wang and T. Xiang, Phys.\ Rev. {\bf B56}, 5061 (1997);
N. Shibata, J. Phys.\ Soc.\ Jpn.\ {\bf 66}, 2221 (1997);
K. Maisinger and U. Schollw\"{o}ck, Phys.\ Rev.\ Lett.\ {\bf 81}, 445 (1998).
\bibitem{Hiei98} Y. Hieida, J. Phys. Soc. Jpn. {\bf 67}, 369 (1998).
\bibitem{Kaul98} M. Kaulke and I. Peschel, Eur. Phys. J. {\bf B5}, 727 (1998).
\bibitem{Kaul99} \"{O}. Legeza, M. Kaulke and I. Peschel,  
Ann. der Physik {\bf 8}, 153 (1999)
\bibitem{Mars97} J. Kondev and J. B. Marston, Nucl.\ Phys.\ {\bf B 497}, 639
(1997); J. B. Marston and Shan-Wen Tsai, Phys.\ Rev.\ Lett. {\bf 82}, 4906
(1999); T. Senthil, J. B. Marston, and M. P. A Fisher, Phys.\ Rev.\ {\bf
B 60}, 4245 (1999); Shan-Wen Tsai and J. B. Marston, 
Ann. der Physik, {\bf 8}, 261 (1999).
\bibitem{Peli86} L. Peliti, J. Phys. A: Math. Gen. {\bf 19}, L365 (1986).
\bibitem{Lee94} B.P. Lee, J. Phys. {\bf A27}, 2633 (1994).
\bibitem{Mend94} I. Jensen and R. Dickman, \pre {\bf 48}, 1710 (1993);
J.F.F. Mendes, R. Dickman, M. Henkel and M.C. Marques, J. Phys. A: Math. Gen.
{\bf 27}, 3019 (1994); M.A. Mun\~oz, G. Grinstein, R. Dickman and R. Livi, 
\prl {\bf 76}, 451 (1996); P. Grassberger, H. Chat\'e and G. Rousseau, 
\pre {\bf 55}, 2488 (1997).
\bibitem{Dick99} R. Dickman, preprint {\tt cond-mat/9909347}
\bibitem{Jans81} H.K. Janssen, Z. Phys. {\bf B42}, 151 (1981); 
P. Grassberger, Z. Phys. {\bf B47}, 365 (1982)
\bibitem{Park95} H. Park and H. Park, Physica {\bf A221}, 97 (1995).
\bibitem{danke} We thank H. Hinrichsen for a useful discussion on that point.
\bibitem{Buli64} R. Burlisch and J. Stoer, Numer. Math. {\bf 6}, 413 (1964);
M. Henkel and G. Sch\"utz, J. Phys. A: Math. Gen. {\bf 21}, 2617 (1988).
\bibitem{Ziff86} R. Ziff, E. Gulari and Y. Barshad, \prl {\bf 56}, 2553 (1986).
\bibitem{Bida89} R. Bidaux, N. Boccara and H. Chat\'e, \pra {\bf 39}, 3094
(1989). 
\bibitem{Dick91} R. Dickman and T. Tom\'e, \pra {\bf 44}, 4833 (1991).
\bibitem{Odor93} G. \'Odor, N. Boccara and G. Szabo, \pre {\bf 48}, 3168 (1993).
\bibitem{Meny95} N. Menyh\'ard and G. \'Odor, J. Phys. {\bf A28}, 4505 (1995).
\bibitem{Odor96} G. \'Odor and A. Szolnoki, \pre {\bf 53}, 2231 (1996).
\bibitem{Oerd99} K. Oerding, F. van Wijland, J.-P. Leroy and H.J. Hilhorst,
J. Stat. Phys. {\bf 99}, 1365 (2000).
\bibitem{Gras84} P. Grassberger, F. Krause and T. von der Twer, 
J. Phys. A: Math. Gen. {\bf 17}, L105 (1984); 
P. Grassberger, J. Phys. A: Math. Gen. {\bf 22}, L1103 (1989).
\bibitem{Meny94} N. Menyh\'ard, J. Phys. A: Math. Gen. {\bf 27}, 6139 (1994).
\bibitem{Inui98} 
N. Inui and A. Yu. Tretyakov, Phys. Rev. Lett. {\bf 80}, 5148 (1998); 
M.C. Marques and J.F.F. Mendes, Eur. Phys. J. {\bf B12}, 123 (1999). 
\bibitem{Hinr97} H. Hinrichsen, \pre {\bf 55}, 219 (1997).
\bibitem{Hinr00a} H. Hinrichsen, {\tt cond-mat/0001177}. 
\bibitem{Odor00} G. \'Odor, Phys. Rev. {\bf E62}, R3027 (2000). 
\bibitem{Barm93} M. Barma, M.D. Grynberg and R.B. Stinchcombe, Phys. Rev. 
Lett. {\bf 70}, 1033 (1993).
\end{references}
\end{document}